# Community-based Behavioral Understanding of Crisis Activity Concerns using Social Media Data: A Study on the 2023 Canadian Wildfires in New York City

**Khondhaker Al Momin**
Ph.D. Student
School of Civil Engineering and Environmental Science
University of Oklahoma
202 W. Boyd St., Norman, OK 73019-1024
Email: momin@ou.edu

**Md Sami Hasnine, Ph.D.**
Assistant Professor
Department of Civil and Environmental Engineering
Howard University
2400 6th St NW, Washington, DC 20059
Email: mdsami.hasnine@howard.edu

**Arif Mohaimin Sadri, Ph.D.**
Assistant Professor
School of Civil Engineering & Environmental Sciences
University of Oklahoma
202 W. Boyd St., Norman, OK 73019-1024
Email: sadri@ou.edu
(Corresponding Author)

** Equal contribution from all authors.

## ABSTRACT

New York City (NYC) topped the global chart for the worst air pollution in June 2023, owing to the wildfire smoke drifting in from Canada. This unprecedented situation caused significant travel disruptions and shifts in traditional activity patterns of NYC residents. This study utilized large-scale social media data to study different crisis activity concerns (i.e., evacuation, staying indoors, shopping, and recreational activities among others) in the emergence of the 2023 Canadian wildfire smoke in NYC. In this regard, one week (June 02 through June 09, 2023) geotagged Twitter data from NYC were retrieved and used in the analysis. The tweets were processed using advanced text classification techniques and later integrated with national databases such as Social Security Administration data, Census, and American Community Survey. Finally, a model has been developed to make community inferences of different activity concerns in a major wildfire. The findings suggest, during wildfires, females are less likely to engage in discussions about evacuation, trips for medical, social, or recreational purposes, and commuting for work, likely influenced by workplaces maintaining operations despite poor air quality. There were also racial disparities in these discussions, with Asians being more likely than Hispanics to discuss evacuation and work commute, and African Americans being less likely to discuss social and recreational activities. Additionally, individuals from low-income neighborhoods and non-higher education students expressed fewer concerns about evacuation. This study provides valuable insights for policymakers, emergency planners, and public health officials, aiding them in formulating targeted communication strategies and equitable emergency response plans.

## BACKGROUND AND MOTIVATION

The COVID-19 pandemic substantially altered people's travel behavior worldwide (*1*) with New York City (NYC) being a central location in the United States. In 2020, compared to the year prior, there was a drastic drop in the number of people using the subway, with a decrease of 91%. Additionally, traffic across bridges and tunnels managed by the Metropolitan Transportation Authority (MTA) also significantly reduced, witnessing a 68% decline (*2*). The implementation of policies and limitations due to COVID-19 caused a stark reduction in the usage of both public transportation and personal vehicles in NYC. The widespread use of technology-based transportation services and modern communication devices, especially smartphones, resulted in a diverse range of travel behaviors during the pandemic, varying by location, time, transport mode, and socioeconomic factors.

Most recently in 2023, Canada experienced the worst-ever wildfire season, with over 400 fires burning across nearly all its regions, devastating over 6.7 million acres and causing around 26,000 people to evacuate (*3*). The resulting smoke has caused a surge in air pollution, triggering air quality alerts across the United States, affecting millions in the Midwest and most of New England. As a result, on Wednesday, June 07, 2023, New York City topped the global chart for the worst air pollution, owing to the wildfire smoke drifting in from Canada. For two days, an orange haze cloaked the city due to a hazardous Air Quality Index (AQI) of 342, affecting all residents and prompting some to wear face masks outdoors (*3*).

This unprecedented situation caused significant travel disruptions and shifts in the traditional activity patterns of NYC residents. For example, despite the city's schools remaining open, outdoor activities have been canceled due to anticipated worsening air quality. Moreover, the smoke has led to the Federal Aviation Administration suspending some flights to LaGuardia Airport, and visibility issues caused delays at the Newark Airport. On the other hand, recreational activities such as major games and musical performances were postponed due to hazardous air quality in their respective cities (*3*). The continual presence of wildfire smoke may suggest increasing toxicity over time, although studies on this subject are limited (*4*). As such, officials from the city have recommended that residents curtail their outdoor activities on Wednesday, with a particular caution for children, the elderly, and individuals with existing respiratory issues who are at a higher risk.
Twitter is a leading microblogging sites in the United States (US), with approximately 200 million daily active users (*5*). Twitter allows users to express their viewpoints, engage in activities, and share ideas through concise 280-character messages called *tweets*. Moreover, geotagged tweets incorporate tweet text, hashtags, and location information, making them akin to check-in data as they reveal the specific posting location of the tweet (*6*). With the rise of Social Media Portals (SMPs) and the growing engagement of people with online media, transportation service providers have been presented with a valuable chance to gather real-time information from SMPs users while minimizing their expenses (*7*).

The predominant data sources employed in the majority of studies related to transportation trends (*8,9*) are surveys (e.g., travel surveys). Studies on evacuation decision-making also primarily rely on surveys of protective action decisions of people from communities affected by major hurricanes or wildfires (*10-18*). Nevertheless, survey data comes with certain limitations, including disparities in data collection methods and availability across different countries, the absence of real-time engagement with respondents, and the high cost and time requirements associated with conducting surveys. In contrast, Social Media Portals (SMPs) have the potential to address these drawbacks by providing a more unified, less intrusive, easily accessible, and comprehensive source of data for transportation studies. Though social media data contains many noises, making it challenging to keep the topical relevance of the research, text classification using machine learning approaches (*19-21*) can improve topical relevancy. Recently, transportation researchers have extensively used social media data (*20-22*), some being in the context of major disasters (*23-29*).

This study utilized large-scale social media data to study different crisis activity concerns (i.e., evacuation, staying indoors, shopping, recreational trips, among others) in the emergence of the 2023 Canadian wildfire smoke in NYC. For the analysis, a dataset comprising approximately one week's worth of Twitter data (from June 2, 2023, to June 9, 2023) has been collected, containing a total of 594,110 tweets originating from NYC. This study has three contributions to the existing literature:

- The study developed a multi-staged text-classification approach using BERT (Bidirectional Encoder Representations from Transformers), a deep learning-based, transformer architecture language model. Using this framework, different crisis activity concerns have been classified from Twitter data.
- Then this study extracted user-level (e.g., gender, race) and geographical-level (e.g., per capita income) attributes from national databases [Social Security Administration (*30*), census (*31*), and American Community Survey (*32*)].
- Finally, this study developed a multinomial logit model to understand the factors affecting these crisis activity concerns. The ultimate goal of this study is to show how Twitter data can be used for evidence-based policy analysis by using advanced text classification and econometric approaches simultaneously.

This study classified wildfire-related tweets into eight distinct categories in accordance with the National Household Travel Survey (NHTS) in the United States. These categories include commuting to work, school trips, shopping and errands, social and recreational activities, medical and dental services, evacuation-related tweets, tweets with other purposes, and tweets related to non-travel or staying at home. By organizing the wildfire tweets into these comprehensive classes, the study aims to gain valuable insights into the diverse aspects of travel and activities impacted by wildfires. This study also provides real-time and geographically specific insights into public travel activities during crisis situations. The use of advanced machine learning techniques and social media data, i.e., Twitter data, can assist policymakers and emergency management in crafting timely, effective responses and advisories, with a special emphasis on the most vulnerable populations. The insights found from this study can help in transportation planning during emergencies and aid in developing policies that consider socioeconomic disparities, thereby making it a critical tool for evidence-based policy analysis.

## DATA COLLECTION AND DESCRIPTION

### *Twitter Data*

This study used the Twitter Academic Application Programming Interface (API) (*33*) for collecting geotagged tweets from the NYC area from June 02, 2023, to June 09, 2023. Twitter academic API can retrieve all tweets that match the given search query through its full-archive search endpoint (*34*). The study collected 0.59M tweets from approximately 66 858 unique users in one week from the NYC area; the spatial distribution of the collected tweets is shown in Figure 1. A significant portion of the tweets were generated in New York County (Manhattan) and Kings County (Brooklyn).

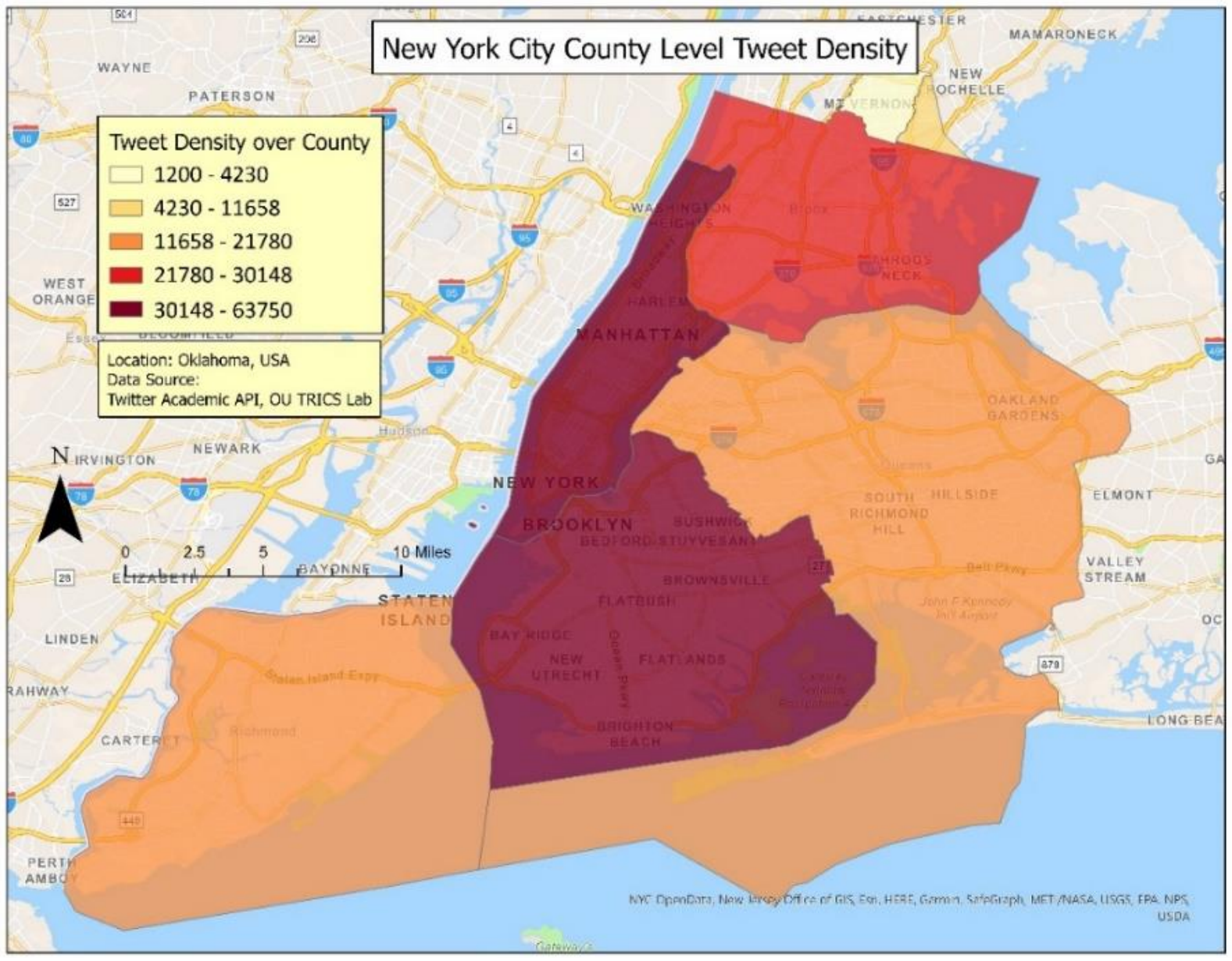

Figure 1: Study Area - New York city county wise tweet distribution

### *United States Social Security Administration Data*

Since its establishment in 1879, the Social Security Administration (SSA) (*30*) has meticulously collected a vast array of first names via social security card applications for births that have taken place within the United States. This wealth of data has proven useful in a variety of fields, particularly when accurate demographic insights are required. This study sought to explore the gender dynamics within the Twitter landscape, and for this purpose, the study turned to the SSA's database. The study leveraged the massive repository of first names available on the SSA website to aid in this endeavor.
As per the records available up to the year 2021, the SSA database contained 63,152 unique male names and 37,212 unique female names. This robust volume of data offers a highly reliable and diverse basis for the training of machine learning models. Specifically, these models can harness this dataset to accurately predict gender based on first names alone, aiding in demographic analysis within social platforms such as Twitter. The reliability of this data, combined with the increasing precision of machine learning algorithms, provides an effective framework for parsing gender demographics from the username on Twitter.

***United States (US) Census Bureau Data***
The US Census Bureau undertakes the annual task of assembling estimates of racial and Hispanic origin proportions across every county within the nation. This endeavor involves the utilization of respondents' surnames as a primary source of data. These estimates are derived from the latest decennial census data and subsequently fine-tuned to factor in changes in the population, including deaths, births, and migratory patterns since the time of the census.
In an effort to streamline this research, the intricacy of race and ethnicity is delineated into four primary groups: Asian, Black, Hispanic, and White. This categorization approach serves to simplify the analysis while still capturing the broad racial and ethnic landscape of the nation. This study collected last names associated with the various racial and ethnic groups from the Census Bureau (*31*). The intention behind this approach was to draw an inference about a Twitter user's race or ethnicity based on their surname. The dataset procured for this study is extensive, encompassing a total of 162,254 unique surnames representative of a multitude of racial and ethnic backgrounds. The study intends to delve into the racial background of Twitter users by utilizing this large dataset. The goal of this study is to gain new insights into the racial and ethnic composition of the Twitter users' using this comprehensive surname dataset.

**Data Preparation**
The study embarked on a multi-faceted process of data preparation. The initial step entailed preprocessing of the Twitter data, which involved meticulous extraction and cleaning of tweet texts, user information, profile description and location data. This stage necessitated the elimination of noise elements such as HTML tags, emojis, and stop words. Furthermore, these collected data underwent a process of tokenization and lemmatization, resulting in smaller, more manageable units conducive to analysis. In an effort to augment the depth of the research, this study integrated demographic and socioeconomic factors with the Twitter data, thereby constructing a rich, multidimensional dataset. To determine the geographical coordinates, latitude and longitude of tweets, the study used reverse geocoding using the Census Geocode Python package. The final refinement of the dataset involved narrowing the scope to include only those tweets that were relevant to wildfire and transportation activity, i.e., commuting to work, school trips, shopping and errands, social and recreational, medical, and dental services, evacuation, other purposes, and non-travel/stay home. This was achieved through the application of relevant filtering techniques as outlined by Kryvasheyeu et al. (*35)*. Upon completion of the relevance filtering process, the study gathered a refined dataset of 10, 258 tweets specifically addressing travel activities during wildfire.

# METHODOLOGY

## Race and Gender Model

The gender-race (GR) model aims to accurately predict gender and race from first names and surnames, respectively. To achieve this, a multi-step process was undertaken in the development of the GR model. Initially, comprehensive data was gathered, consisting of over 100,000 first names from the SSA database and more than 160,000 surnames from the US Census Bureau. Once the data was collected, meticulous preprocessing was performed to ensure its appropriateness for further analysis. The last names were extracted from the source file and organized into a structured data frame. To establish crucial patterns, a specialized function was devised to calculate the occurrences of each letter within a name. These letter counts played a pivotal role in the creation of an alphabet matrix, which was enriched with additional race and gender labels corresponding to each name.

The subsequent stage of the process involved meticulous feature engineering, resulting in the generation of a model-ready dataset. Leveraging the alphabet matrix, the letter counts were converted into concatenated strings, which were then paired with race labels spanning four distinct categories: Asian, Black, Hispanic, and White. In a similar fashion, each string was associated with one of two gender labels: Male or Female. Through this systematic approach, the gender-race (GR) model was developed with great care and precision, laying the foundation for accurate gender and race predictions based on first and last names.

### *Model Training and Evaluation:*

The alphabet matrix was segregated into a feature matrix, X, incorporating the letter frequencies, and a target vector, y, incorporating the race and gender identifiers. The dataset was subsequently partitioned into training and validation subsets in compliance with a 70-30 train-test ratio. This division provided an adequate quantity of data for model training while assuring rigorous assessment. A variety of machine learning algorithms, including Random Forest (RF), Decision Tree (DT), K-Nearest Neighbor (KNN), Support Vector Machine (SVM), and Naïve Bayes (NB), were employed to instruct the models on the training data to predict race and gender identifiers from given letter frequencies. The model's accuracy was gauged on the validation set, and out-of-bag accuracy was computed to counteract the possibility of overfitting. Further, a ten-fold cross-validation was carried out to approximate the proficiency of the machine learning models on unseen data and to enhance resilience against the randomness of the training and validation data partition.

The efficacy of the models was appraised by employing a range of metrics. The accuracy score furnished an insight into the model's proficiency in predicting gender and race identifiers based on first name and surnames, respectively. To illustrate the quantity of correct and erroneous predictions for each gender and racial category, a confusion matrix was concocted. Furthermore, a classification report was generated, encompassing precision, recall, F1-score, and support for each gender and racial category. These metrics offered a holistic comprehension of the model's performance across diverse categories and facilitated the identification of potential areas of enhancement.

**Bidirectional Encoder Representations from Transformers Model**

Bidirectional Encoder Representations from Transformers (BERT) model is a groundbreaking model in the field of Natural Language Processing (NLP), primarily due to its bi-directional nature and attention mechanisms. The model undergoes two main phases: pre-training and fine-tuning. Pre-training involves learning language representation by training on a large corpus of text such as Wikipedia, using unsupervised tasks like Masked Language Model (MLM) and Next Sentence Prediction (NSP). Once pre-trained, BERT is then fine-tuned on a task-specific dataset to cater to a particular NLP task, like text classification.

The input to BERT is a sequence of tokens, which are essentially words in a sentence mapped to unique integer identifiers. The sequence is processed with the *WordPiece* tokenization scheme. BERT also employs special tokens such as [*CLS*] for classification tasks and uses the final hidden state of this token as the aggregate sequence representation. An important aspect of BERT is its ability to capture word positions and sentence boundaries. For this purpose, it uses Positional and Segment Embeddings. These embeddings are combined with the token embeddings to form the final input representation.

The core of the BERT model consists of a multi-layer bidirectional Transformer encoder. A pivotal component of this encoder is the self-attention mechanism. Given an input sequence of token embeddings $X = \{x_1, x_2, ..., x_n\}$, for each token $x_i$, a set of Query (Q), Key (K), and Value (V) vectors are computed:

$Q_i = W_Q x_i$ *(Query transformation)*
$K_i = W_K x_i$ *(Key transformation)*
$Vi = W_V x_i$ *(Value transformation)*

Where $W_Q$, $W_K$, and $W_V$ are the learned weight matrices. The attention score (S) between two tokens, i and j, is computed using their respective query and key vectors. It's normalized by the square root of the key vector dimension $d_K$ to prevent large dot product values:

$$S_{ij} = \left(\frac{Q_i.K_j}{\sqrt{d_k}}\right) \quad (1)$$

This attention score is then passed through a *softmax* function to determine the weights for the attention mechanism:

$$\alpha_{ij} = \text{SoftMax}\,(S_{ij}) \quad (2)$$

The output (O) for each token is then computed as the sum of all value vectors, weighted by the attention scores:

$$O_i = \sum_j \alpha ij \cdot V_j \quad (3)$$

In the BERT model, these computations are performed bidirectionally, allowing each token to attend to all other tokens, irrespective of their position. Moreover, the model uses multiple sets of Q, K, and V vectors for each token, which provides a richer representation of the input. For classification tasks, the output of the [*CLS*] token from the last transformer layer serves as the sequence representation, which is then fed into a task-specific classifier. During fine-tuning, the classifier and BERT model weights are jointly trained to align the model to the specific task.

**Multinomial Logit (MNL) model**

A multinomial logit (MNL) model was developed to understand the public concern on different activity types (evacuation, staying at home, work commute, trips to school/church, shopping/errands, social/recreational trips, and other purposes).

Derived from consumer economics theory developed by McFadden (*36)*, the MNL model stands as the most widely utilized type of random utility model. Within this model, an individual, denoted as $i$, selects a single choice from a set of discrete alternatives by assessing the features $X$ associated with each alternative in a utility maximization process. Ultimately, the individual $i$ selects the alternative $m$ that yields the highest utility:

$$U_{im} > U_{ik} \, ; m \neq k \tag{4}$$

The concept of utility can be divided into two components: the observed utility, $V_{im}$ and the unobserved utility $\varepsilon_{im}$, which together represent the complete utility experienced by a person. To assess the comprehensive utility, $V_{im}$ encompasses two sets of attributes: the covariates linked to both the individual and the alternative, $X_{im}$, and the decision-maker characteristics, *Si (37)*

The observed utility (V) is a numerical representation derived from a linear function of the attributes used in the analysis. This value serves to quantify the level of popularity or preference associated with a particular option within the limitations set by the model's specifications.

$$V_{im} = V(X_{im}, S_i) \tag{5}$$

The MNL model is formulated based on the explicit assumption that each unobserved term, $\varepsilon_{im}$, follows an Independent and Identically Distributed (IID) extreme value distribution, specifically the Gumbel or type 1 extreme value distribution. By leveraging this assumption, we can determine the likelihood of an individual $i$ selecting alternative $j$ by solving the corresponding mathematical formula:

$$P_{ij} = \frac{e^{\beta' X_{im}}}{\sum_{k=1}^{M} e^{\beta' X_{ik}}} \tag{6}$$

Where $X_{ik}$ represents the vector of observed explanatory variables associated with choosing a specific alternative, and $\beta'$ denotes the parameter corresponding to the observed utility. For a deeper understanding of the technical aspects related to logit models within the discrete choice methodology, further exploration of relevant literature can be found in reference (*37*).

## RESULTS

### Gender-Race model

In this research, the authors collected first names (used for gender prediction) from the Social Security Administration database, with a total of 100,364 unique first names, and also gathered surnames (used for race prediction) from the United States Census Bureau, amounting to 162,254 unique last names. These data were then employed to train a wide array of machine learning classifiers, such as Random Forest (RF), Decision Tree (DT), K-Nearest Neighbor (KNN), Support Vector Machine (SVM), and Naïve Bayes (NB) models. The efficacy of these machine learning models in correctly identifying gender is depicted in Figure 2.

Upon examining performance metrics, the Random Forest (RF) model stands out as the most effective for predicting gender (either male or female). The RF model boasts a remarkable average accuracy of 0.84 and demonstrates strength in precision, recall, and F1-score. This suggests a well-balanced strategy for dealing with both false positives and false negatives. The evidence strongly attests to the RF model's ability to accurately classify instances and maintain an optimal balance between precision and recall. Thus, the study used the RF model for predicting gender from the collected Twitter usernames.

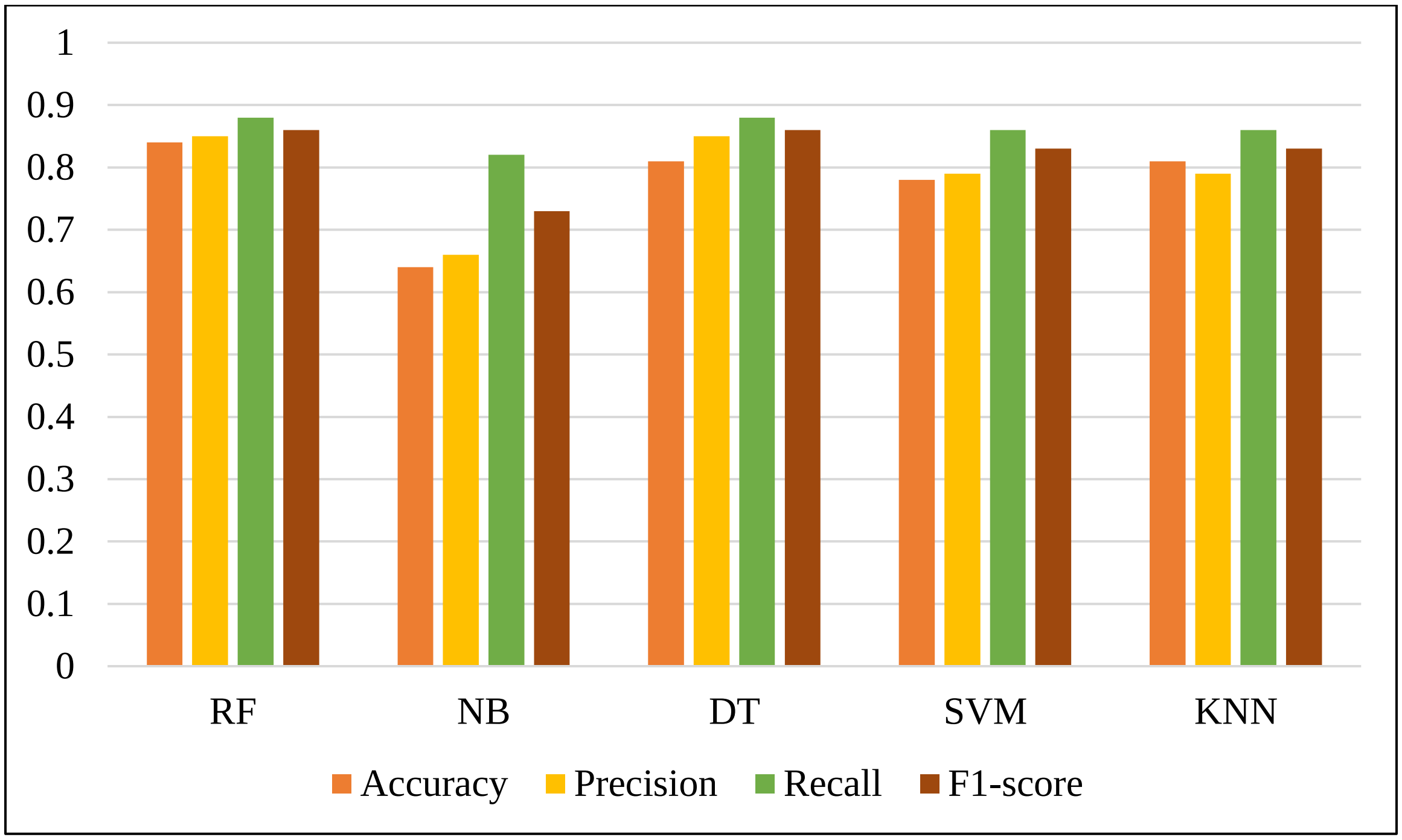

Figure 2: Gender Prediction Model Performance values

This research found SVM model is the most apt for predicting race among the four groups: Asian, Black, Hispanic, and White, as illustrated by the performance metrics (see Figure 3). The SVM model delivered the highest average accuracy score of 0.84, in addition to showcasing outstanding precision, recall, and F1 score. This points to a substantial equilibrium between the volume of accurate predictions and the proportion of true positive predictions. This sort of balance is essential when dealing with multiclass predictions, ensuring the model's solid performance across all categories. Given these performance metrics and the problem's multiclass nature, the SVM model is deemed most effective for race prediction.

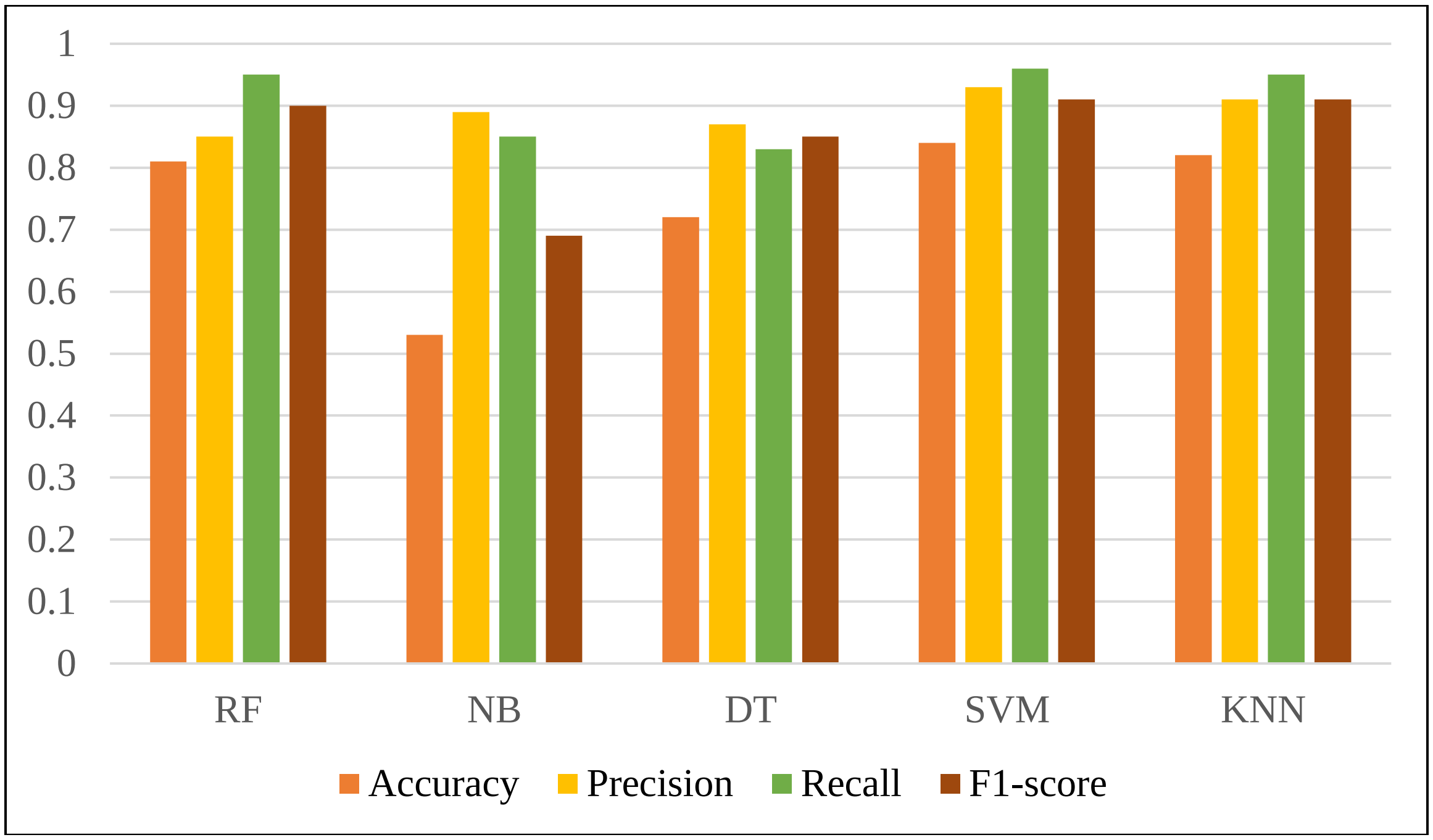

Figure 3: Race Prediction Model Performance

**BERT Text Classification Result**

Based on the findings of the National Household Travel Survey (NHTS) in the United States, this study has classified wildfire-related tweets into eight distinct categories. These categories include commuting to work, school trips, shopping and errands, social and recreational activities, medical and dental services, evacuation-related tweets, tweets with other purposes, and tweets related to non-travel or staying at home. After relevant filtering, the study got 10258 tweets specifically expressing concern about travel activities during the wildfire. The authors labeled 2500 tweets in these eight distinct categories for training and testing the BERT model. Equipped with this meticulously labeled data set, the study was able to effectively harness the BERT model, utilizing its advanced natural language processing prowess. This rigorous training methodology primed the BERT model to provide detailed and contextually precise analyses across the wide range of themes in the dataset.

$$Precision\ (for\ a\ given\ class) = \frac{True\ Positives}{True\ Positives + False\ Positives} \quad (7)$$

$$Recall\ (for\ a\ given\ class) = \frac{True\ Positives}{True\ Positives + False\ Negatves} \quad (8)$$

$$F1\ score\ (for\ a\ given\ class) = \frac{2 \times Precision \times Recall}{Precision + Recall} \quad (9)$$

Metrics such as precision (Eqn. 7), recall (Eqn. 8) and F1 score (Eqn. 9) are typically used to evaluate the effectiveness of a machine learning model, particularly for classification tasks *(38)*.

The BERT model, diligently trained using the labeled dataset of this study, has shown outstanding performance, flaunting a remarkable F1 Score of 0.965 with a Precision value of 0.974, securing the perfection of its positive predictions. Additionally, it showcased an extraordinary Recall of 0.986, adeptly identifying an extensive proportion of relevant instances. Such outstanding figures underscore the efficiency and dependability of the BERT model in dealing with the tasks posed in this research.

Figure 4 depicts the text classification results in both percentage and actual value in a pie chart. This categorization allows for a clear and organized representation of the data, allowing for a more in-depth understanding of the relationships and patterns within the text categories mentioned.

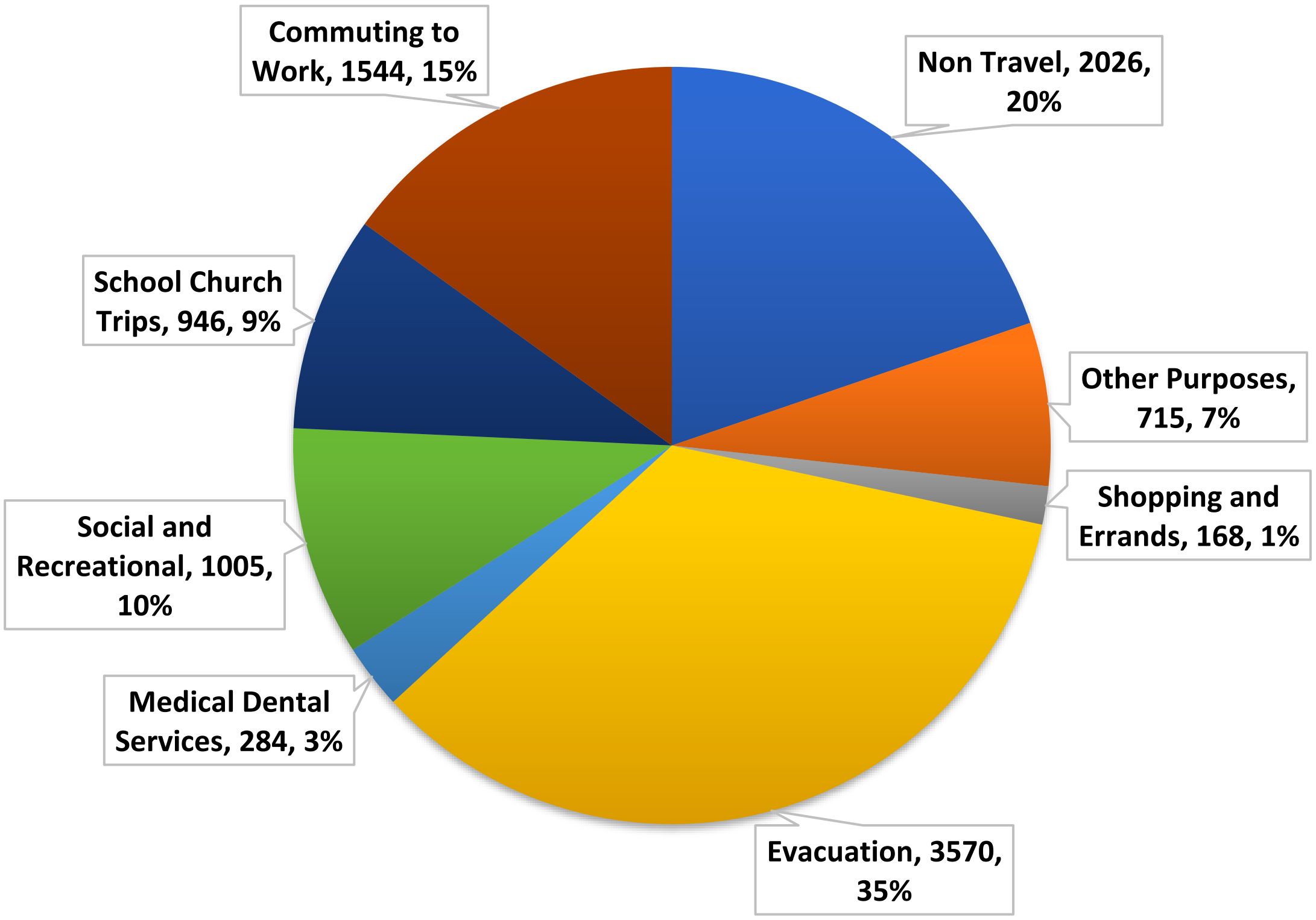

Figure 4: BERT Text classification outcome

**Sentiment Analysis of Travel Activities**

Figure 5 visually demonstrates a heatmap that provides an intriguing look into how sentiments are scattered across diverse travel activities. A standout sighting is the strikingly high volume of tweets related to *Evacuation,* which considerably overshadows other categories. This implies an intense discussion and unease concerning situations involving evacuations. It is also observable that people expressed concern about the Non-travel/staying at home and Commuting to work.

The sentiment analysis goes a step further to underscore an immense amount of negative sentiment across almost all travel categories. This might hint at an overall trend of discontentment or critique associated with these travel actions. Notably, while the *Evacuation, Non-Travel and Commuting* *to Work* category garnered the most tweets, it also faced a substantial brunt of negative sentiments. This might signify public frustration or anxiety surrounding evacuation decisions, commuting to work, and staying at home during the wildfire. On the flip side, the *Other purposes, shopping and Errands* and *medical and dental services* categories logged fewer tweets. This could insinuate these activities were either less disputed or simply not as frequently discussed as other categories. Comparing the distribution of neutral and negative sentiments, it's evident that tweets expressing positive sentiments are less frequent across all categories. This indicates that users are more prone to tweet about these topics in response to negative experiences or neutral observations rather than positive ones. For instance, tweets regarding *Non-Travel, School Church Trips*, and *Social and Recreational* have a substantial amount of neutral tweets. This lack of strongly positive or negative sentiment could imply a general ambivalence about these travel activities, with individuals neither lauding them nor airing serious grievances, leading to a preponderance of neutral sentiment tweets. These insights can be invaluable for policymakers, urban planners, and transport agencies, providing a detailed picture of public sentiment and key areas of concern concerning various travel activities.

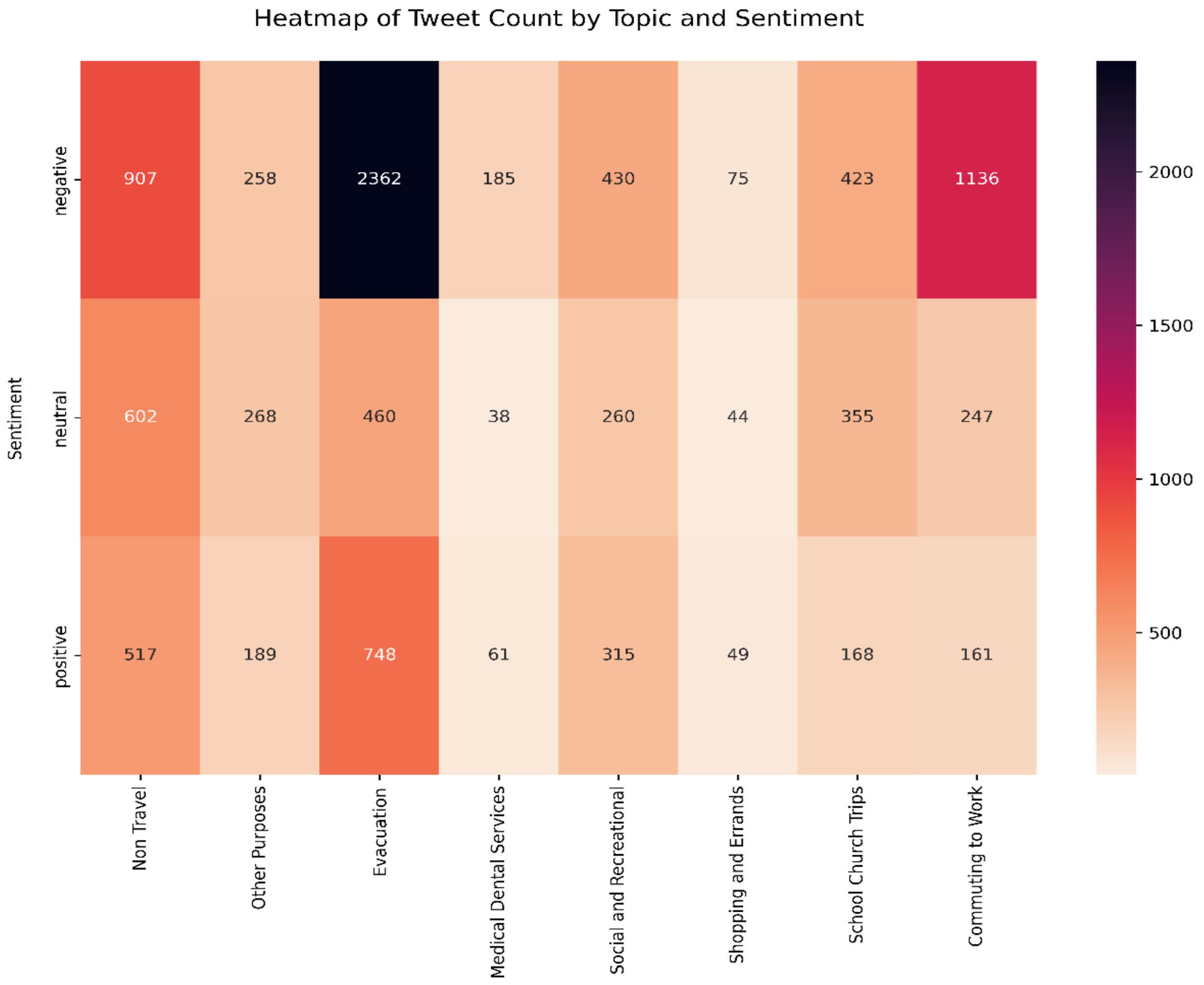

Figure 5: Sentiment Analysis of Travel Activities

**Model Estimation Results**

The table shows various factors affecting the individuals' activity type in NYC during wildfires in Canada. The sample size is 10,258, and the goodness-of-fit against the null model is 0.25, which is a reasonable fit. Most of the reported parameters are statistically significant within 95% confidence intervals.

The model includes eight alternatives, such as commuting to work, school trips, shopping and errands, social and recreational, medical and dental services, evacuation, other purposes, and non-travel/stay home. A total of seven alternative-specific constants are estimated in the model. All alternative-specific constants are statistically significant, and the commuting to work alternative is considered as the reference alternative.

It is found that females are less likely to discuss about evacuation during wildfires. Individuals who lived in NYC in June 2023 were not directly impacted by this wildfire. However, it was advised by the NYC government officials that individuals should stay at home to avoid wildfire-related air pollution. In addition, it was reported that wildfire-related air pollution caused health issues for healthy adults, the elderly, and children. As a result, many NYC residents expressed more concern about staying at home rather than evacuating.

It is also found that females are less likely to discuss any trips related to medical and dental treatment. Similar trends have been found in terms of social and recreational trips. The model results reveal that females are less sensitive to commuting to work during wildfires. The rationale for this behavior is that many companies in NYC still operated their regular businesses during the wildfire. Therefore, commuters who go to work have no choice but to commute while the air quality is severely poor.

This model also tested various categories of race and how different individuals' activity evolved during the wildfire. It is found that Asians are more likely to discuss wildfire evacuation than Hispanics. It is also found that Asians are more likely to express concern about work during wildfires. African American are less likely to discuss trips related to social and recreational activities. Hispanic individuals revealed different activity patterns than African American. The model also incorporated various aggregate sociodemographic attributes. Areas where low-income group individuals live and residents that are not higher educational students are less likely to express concern about evacuation during a wildfire.

**Table 1: Model Estimation Results**

| | |
|---|---|
| Number of observations | 10,258 |
| Loglikelihood of estimated model | -16,007.199 |
| Loglikelihood of null model | -21,330.896 |
| Rho-squared against null model | 0.250 |

| **Variable** | **Utility** | **Parameter Estimate** | **t-stat** |
|---|---|---|---|
| Alternative Specific Constant (ASC) | School Trips | -0.435 | -5.766 |
| | Shopping and Errands | -1.623 | -17.476 |
| | Social and Recreational | 0.277 | 4.438 |
| | Medical and Dental Services | -1.440 | -13.733 |
| | Evacuation | 1.843 | 35.792 |
| | Other Purposes | -0.281 | -3.979 |
| | Non-Travel | 0.809 | 14.287 |
| Female | Commuting to Work | -0.084 | -1.000 |
| | School Trips | -0.084 | -0.857 |
| | Social and Recreational | -0.311 | -3.915 |
| | Medical and Dental Services | -0.228 | -1.440 |
| | Evacuation | -0.177 | -3.328 |
| | Other Purposes | -0.127 | -1.361 |
| Race: Asian | Commuting to Work | 0.338 | 2.630 |
| | School Trips | 0.206 | 1.287 |
| | Evacuation | -0.120 | -1.459 |
| Race: Hispanic | Social and Recreational | 0.339 | 1.636 |
| | Evacuation | -0.106 | -0.738 |
| Race: African American | Social and Recreational | -0.368 | -1.078 |
| Low-income group and high percent of residents that are not higher education students (Dummy variable 0 or 1) | Evacuation, Social and recreational | -0.061 | -0.227 |

## CONCLUSIONS

In recent years, transportation researchers used SMPs extensively to address a wide range of problems, including travel demand forecasting, activity pattern modeling, transit service evaluation, traffic incident analysis, and disaster management, among other areas. However, there remains a vast potential to further investigate how such data can enrich our comprehension of public perception and attitudes towards transportation trends and activity patterns, especially during a major crisis situation. As such, this research aims to introduce a new methodology that can categorize social media posts (i.e., tweets) into different activity types integrating social media user attributes with SSA and US Census data. Additionally, this research conducted an econometric analysis aiming to demonstrate the influence of various socioeconomic and demographic factors on travel activity trends. To achieve this, this study integrated Twitter data with a customized selection of socioeconomic variables obtained from the ACS (*32*).

The multinomial logit model was developed to perform an econometric analysis investigating the popularity of different travel choices in SMPs in the emergence of the 2023 Canadian wildfire impacting the New York City (NYC). The model is based on geotagged tweets collected from NYC between June 02 through June 09 in 2023, using Twitter Academic API. Some key findings from the study include:

- *Females are less likely to discuss about evacuation during wildfires.*
- *Females are also less likely to discuss about trips for medical, dental, social, and recreational purposes during wildfires.*
- *Females were less inclined to discuss commute for work during wildfires, possibly due to employers continuing their operations amidst the wildfires, necessitating commuting despite poor air quality.*
- *NYC residents in June 2023 weren't directly impacted by wildfires, leading many NYC residents to discuss about staying at home.*
- *Racial differences in response to wildfires: Asians were more likely to discuss about evacuation and work commute during wildfires than Hispanics, while African Americans were less likely to engage in discussion related to social and recreational activities.*
- *People from low-income neighborhoods and non-higher educational students were less likely to express concern about evacuation during a wildfire.*

Despite making significant contributions to literature, this study is not beyond limitations. Future studies should eliminate tweets from bots (*39*) following different methods available in the literature (*40,41*). Future studies should consider conducting a survey among Twitter users, incorporating other national databases (e.g., National Household Travel Survey) to improve the model. To make population inferences, future studies could use different sampling techniques by stratifying the datasets based on different sociodemographic factors.

The study showed that Twitter can be used as an effective source to capture travel-related concerns, integrate geotagged social media data with sociodemographic attributes using national databases, and predict people's attitudes towards travel activity types at any geographical scale. The following are some potential applications of the study: (**i**) capturing different travel-related signals from SMPs with high topical relevancy; (**ii**) incorporating existing national databases to investigate and model community mobility behavior at a different level of resolution in the emergency period (e.g., hurricane, pandemic) within a short period before conducting any survey; (**iii**) identifying and forecasting spatial diversity of various travel-related needs and concerns through social media channels; (**iv**) developing new policies aimed at satisfying the diverse mobility requirements at across different locations; (**v**) design and implement more effective approaches to boost public engagement and satisfaction related to various transportation choices.

## ACKNOWLEDGMENTS

The material presented in this paper is based on work supported by the National Science Foundation under Grants No. OIA-1946093 and SCC-PG- 2229439. Any opinions, findings, and conclusions or recommendations expressed in this paper are those of the authors and do not necessarily reflect the views of the National Science Foundation.

## AUTHOR CONTRIBUTIONS

Khondhaker Al Momin: Data collection, Methodology, Formal analysis, Draft preparation and Review & editing. Md Sami Hasnine: Formal analysis, Draft preparation, and Review & editing. Arif Mohaimin Sadri: Conceptualization, Draft preparation, Methodology, Supervision, and Review & editing. All authors contributed equally to this paper. All authors reviewed the results and approved the final version of the manuscript.